\documentclass[runningheads]{llncs}
\usepackage[T1]{fontenc}
\usepackage{graphicx}
\usepackage{xcolor}
\usepackage[urlcolor=blue]{hyperref}      
\hypersetup{
    colorlinks = true,                    
    citecolor = {blue},
    linkcolor = {purple},
           }

\usepackage{amsmath}
\usepackage{amssymb}
\usepackage{bm}

\usepackage{subfig}

\begin{document}
\title{Neural self-organization for muscle-driven robots
}
\titlerunning{Robots with muscles}
%
\author{Elias Fischer$^*$ \and Bulcs\'u S\'andor$^\dagger$ \and
Claudius Gros$^*$
}
\authorrunning{E.\ Fischer et al.
}
%
\institute{
$^*$Institute for Theoretical Physics,
Goethe University Frankfurt a.M., Germany
\\
$^\dagger$Department of Physics, Babes-Bolyai University, Cluj-Napoca, Romania
}
\maketitle              
\begin{abstract} 
We present self-organizing control principles 
for simulated robots actuated by synthetic muscles. 
Muscles correspond to linear motors exerting force 
only when contracting, but not when expanding, with 
joints being actuated by pairs of antagonistic muscles. 
Individually, muscles are connected to a controller 
composed of a single neuron with a dynamical threshold
that generates target positions for the respective
muscle. A stable limit cycle is generated when the
embodied feedback loop is closed, giving rise to
regular locomotive patterns. In the absence of direct
couplings between neurons, we show that force-mediated
intra- and inter-leg couplings between muscles suffice
to generate stable gaits.

\end{abstract}
%

\section{Muscle-driven robots}

A substantial effort is devoted to the development
of robotic artificial muscles \cite{zhang2019robotic},
with possible applications ranging from interactive soft 
robotics \cite{wang2021recent} to the re-creation of
human walking via compliant legs \cite{geyer2006compliant}.
In comparison, only a somewhat limited number of studies have
been devoted to the study of robotic control principles
for synthetic muscles \cite{geyer2010muscle,mohseni2022unified}.
Here we examine control principles based on
embodied self-organization that have been
developed previously for robots driven by
rotating actuators (motors) 
\cite{sandor2018kick,kubandt2019embodied}.
For pairs of antagonistic muscles that are controlled
independently, viz without cross-control, we find 
spontaneous anti-synchronization due to the 
indirect coupling via the moving limb.
Our studies are carried out using
\href{http://www.cyberbotics.com}{Webots}, 
an open-source mobile robot simulation software 
developed by Cyberbotics Ltd \cite{Webots}.

\begin{figure}[t]
\centering
\subfloat{\includegraphics[width=0.4\textwidth]{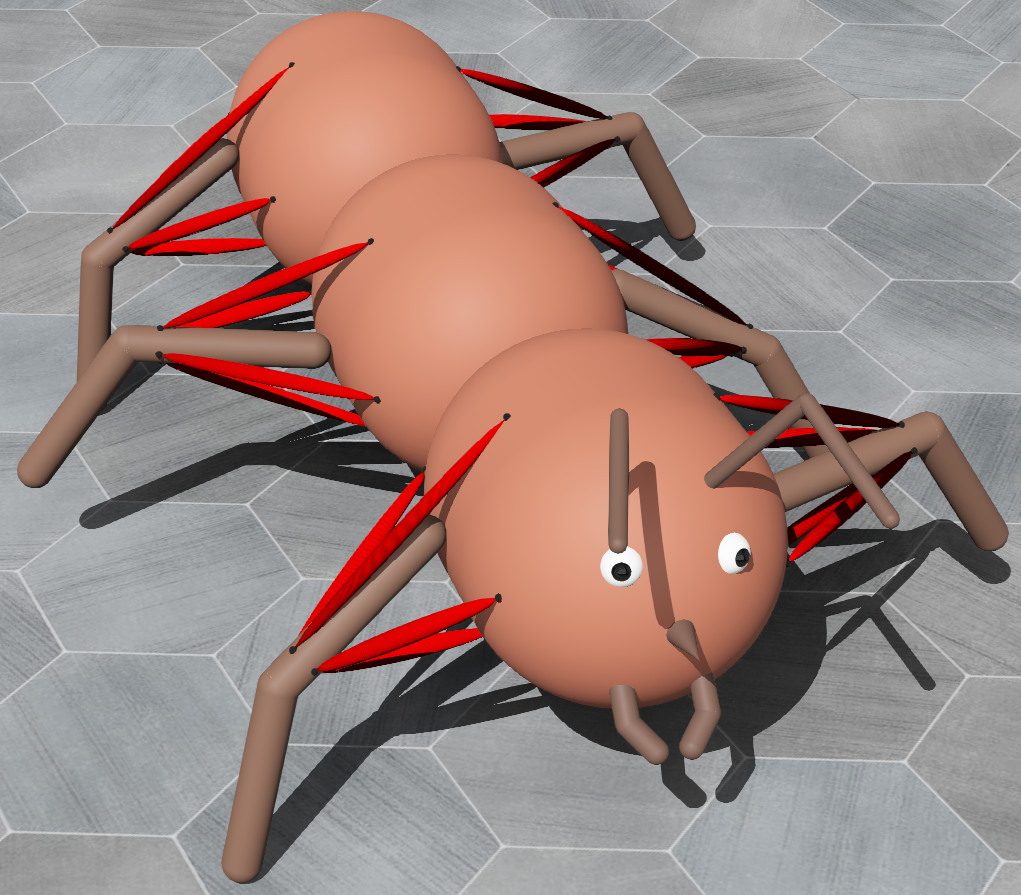}}
\hspace{4ex}
\subfloat{\includegraphics[width=0.35\textwidth]{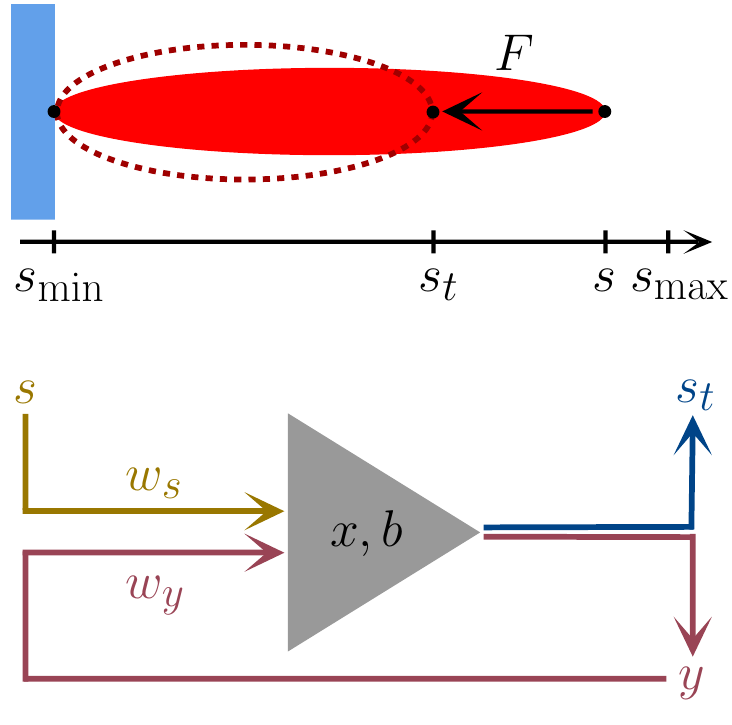}}
\caption{{\bf Left:} Six-legged robot 
driven by 24 muscles. Each leg is controlled 
by two pairs of antagonistic muscles, enabling 
movement both in up-down and forwards-backwards 
direction. Simulations were performed using 
the Webots open-source robot simulation 
software by Cyberbotics Ltd \cite{Webots}.
{\bf Right:} 
Schematics of the single neuron controller. 
The neuron takes the current actuator position 
$s$ and its own activation $y$ as inputs, 
weighted respectively with synaptic weights 
$w_s$ and $w_y$. The target position
$s_t$ determines via (\ref{force_s_t}) the 
actuating force $F$
[\href{https://doi.org/10.6084/m9.figshare.23703399}{Link} to the video].}
\label{fig:Webots_legs}
\end{figure}

The core processing unit of our controller 
is a single neuron with membrane potential
$x(t)$ and a variable threshold $b(t)$. 
The neuron receives two types of 
inputs via constant synaptic weights, 
$w_s$ and $w_y$, as illustrated in
Fig.\,\ref{fig:Webots_legs}.
The first, $w_s$ transmits information 
about the current status $s=s(t)$ of 
the actuator, with the second, $w_y$, corresponding to an excitatory 
self-coupling:
\begin{align}
\label{dot_x}
\tau_x \dot{x} &= -x + w_s s_{\rm rel} + w_y y,
\quad\qquad  
s_{\rm rel} =  \frac{s - s_{\rm min}}{s_{\rm max}-s_{\rm min}},
\\[0.1ex]
\tau_b \dot{b} &= y - y_b, 
\quad\qquad  
y = \frac{1}{1+\mathrm{e}^{a(b-x)}}\,,
\label{dot_b}
\end{align}
where the neuronal activity $y\in[0,1]$ 
is determined by a sigmoidal with gain 
$a$ and threshold $b$. The time
constants for the evolutions of 
membrane potential and threshold are respectively $\tau_x$ and $\tau_b$.
The position $s$ of the actuator is bounded by 
physical constraints, such that $s\in[s_{\rm min},s_{\rm max}]$.
Using the relative position $s_{\rm rel}\in[0,1]$ 
as an input to the membrane potential, 
as done in (\ref{dot_x}), allows to 
directly compare the sizes
of $w_s$ and $w_y$. Entering 
(\ref{dot_b}) is the desired
steady-state value $y_b$ for 
the neural activity $y$. It is 
reached however only if activities 
would cease altogether.

The one-neuron controller acts by generating 
a target position 
$s_t\in[s_{\rm min},s_{\rm max}]$ 
for the actuator, which in turn is 
translated to a force $F$ via
\begin{equation}
F = -\gamma \dot{s} + K_s\, \frac{s_t - s}{s_{\rm max}-s_{\rm min}}, 
\quad\qquad 
s_t = s_{\rm min} + (s_{\rm max} - 
s_{\rm min})y\,,
\label{force_s_t}
\end{equation}
where $K_s$ is the coefficient for 
proportional control and $\gamma$ a
phenomenological damping constant.
The results presented are for critical
damping.
We assume with (\ref{force_s_t}), that 
the target position $s_t$ for the actuator 
is directly proportional to the neuronal
activity $y=y(t)$. As a result, 
one has a sensori-motor feedback loop
\cite{sandor2018kick,kubandt2019embodied},
with the actuator trying to reach a 
continuously updated target position.
Biologically, muscles may exert 
force only when contracting, but not 
when expanding. This corresponds to 
the substitution
%
$F\to F\,\big[1-\theta(F)\big]$,
%
where we use the Heaviside step function $\theta(x)$ to set the force to zero when
$s_t>s$, viz when the length $s$ would be increased.

\begin{figure}[t]
\centering
\includegraphics[width=0.9\textwidth]{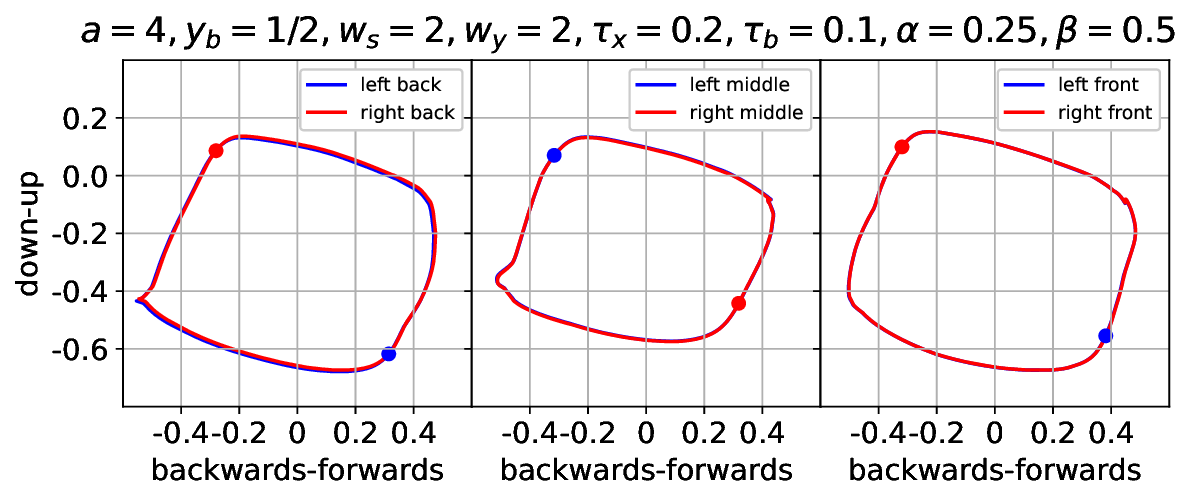}
\caption{The angle (in radians) of the legs 
of the robot shown in Fig.\ref{fig:Webots_legs}, 
viewed from the left side of the robot walking 
to the right after the initial synchronization 
phase. The dots show the position of the respective 
leg at the last time step, showcasing a tripod gait 
with the middle legs being in opposite phases to the 
front and back legs. The blue trajectory of the left 
legs can hardly be seen because the left/right 
trajectories align almost perfectly.}
\label{fig:results}
\end{figure}

\smallskip
\noindent{\bf Attractoring.\ }
The autonomous system, attained by
setting $w_s=0$ in (\ref{dot_x}),
shows a super-critical Hopf transition 
at
\begin{equation}
w_y = \frac{4}{a} + \frac{\tau_x}{\tau_b}\,,
\label{Hopf}
\end{equation}
which holds for $y_b=1/2$. When $w_y$ 
and/or $a$ is large, the system 
oscillates spontaneously, acting as a 
central pattern generator (CPG). In this regime,
the additional feedback $w_s s_{\rm rel}$ 
corresponds to a modulator. Here we 
concentrate on the case that the 
isolated neuron does not oscillate
on its own, viz that $w_y$ and/or $a$ 
is too small for (\ref{Hopf}) to be 
fulfilled. Locomotion is generated
consequently only when the feedback 
from the actuator is strong enough for 
an embodied limit cycle to emerge. 
We call this regime `attractoring', 
which has been found to allow for increased 
behavioural flexibility \cite{sandor2018kick}.
Locomotion is embodied in the sense 
that the phase space of the resulting 
limit cycle contains the degrees of 
freedom  of the body in addition to 
$x(t)$ and $b(t)$. 
We note in this context that it is 
important to use force signals for 
both real-world and simulated
actuators, as the respective default 
PID controllers  tend to be stiff.

\smallskip
\noindent{\bf Force mediated inter-muscle coupling.}
The desired movement for a leg with two pairs 
of antagonstic muscles (up-down; left-right)
is up-forwards-down-backwards. For this we
expand (\ref{force_s_t}) as
\begin{equation}
 s_{t,1} = s_{\rm min} + (s_{\rm max} - s_{\rm min})\cdot 
((1-\alpha)y_1 + \alpha y_2), 
\qquad \alpha\in[0,1]\,,
\label{muscle_coupling}
\end{equation}
which corresponds to an embodied coupling via
force superposition. The activity $y_2$ of a 
second neuron of the same leg influences the 
target position (and hence the force) generated 
by the first neuron, but not the first neuron directly. 
The order of coupling between the 
four muscles of a single leg is taken to be circular.
The same principle is used for (indirect) inter-leg
coupling,
\begin{equation}
s_{t,1} = s_{\rm min} + (s_{\rm max} - s_{\rm min})\cdot 
((1-\alpha - \beta)y_1 + \alpha y_2 + \beta y_3), 
\qquad \alpha + \beta \leq 1\,,
\end{equation}
where $y_3$ is now the activity of a neuron from
another leg. For the six-legged robot shown 
in Fig~\ref{fig:Webots_legs}, the 
contralateral pairs of 
legs are coupled via the 
up-down muscles for producing steps, 
while the inter-leg phase blocking 
is mediated solely via the upper muscles.
We call this coupling principle 
'force-mediated' coupling.

\section{Results}

For parameters in the attractoring
regime, we present in 
Fig.\,\ref{fig:results} the time
evolution of the positions of the six
legs. One observes a stable tripod gait 
[\href{https://doi.org/10.6084/m9.figshare.23703399}{Link} to the video], 
which emerges without the direct 
coupling of the controlling neurons. 
A conceptually similar result has also been 
achieved by using pressure sensors 
and motors \cite{owaki2013simple}, 
albeit relying on CPGs for controlling 
the individual legs.
Note that here oscillations would not 
be generated without feedback from the 
body and no forces are exerted
when the muscles relax, so in this sense 
the locomotion is fully self-organized.


%
%
%


%
\end{document}